\documentstyle[twocolumn]{article}
\newcommand{\kms}{\rm \,km\, s^{-1}}
\newcommand{\co}{{\rm ^{12}CO} (J=1-0)}
\newcommand{\vlsr}{V_{\rm LSR}}
\newcommand{\ico}{I_{\rm CO}}

\newcommand{\Deg}{^\circ}
\newcommand{\arcsec}{''}
\newcommand{\arcmin}{'}
\newcommand{\arcdeg}{^\circ}
\def\bibitem{\hangindent 1pc \noindent}
\def\kmps{km s$^{-1}$}

\begin{document}
\title{The Virgo High-Resolution CO Survey I. \\ -- CO Atlas --}
\author{
   Y. Sofue$^1$,
   J. Koda$^{1,2,3}$,
   H. Nakanishi$^1$,
   S. Onodera$^1$,
   K. Kohno$^1$,\\
   A. Tomita$^4$,
   and
   S. K. Okumura$^3$ \\
\\
{1. Institute of Astronomy, University of Tokyo, Mitaka, Tokyo 181-0015}\\
{2. National Astronomical Observatory, Mitaka, Tokyo 181-8588}\\
{3. Nobeyama Radio Observatory, Minamimaki-mura, Nagano 384-1305}\\
{4. Faculty of Education, Wakayama University, Wakayama 640-8510}\\
\\
{Email: sofue@ioa.s.u-tokyo.ac.jp}
}

\date{ }

\maketitle

\begin{abstract}
We present the results of the Virgo high-resolution CO survey (ViCS)
obtained with the Nobeyama Millimeter-wave Array (NMA).
This survey was made in the course of a long-term project at Nobeyama
from 1999 December through 2002 April.
The objects were selected from Virgo cluster members, considering
CO richness from single dish flux, mild inclination, and lack of
strong tidal perturbations.
The central $1\arcmin$ regions ($\sim 4.7$ kpc) of 15 spiral galaxies
were observed with resolutions of $2-5\arcsec$ and $10-20\kms$, and
sensitivities of $\sim 20\,{\rm mJy \, beam^{-1}}$ for a $10\kms$ channel.
The objects lie at the same distance of the Virgo cluster (16.1 Mpc), which
is advantageous for comparisons among individual galaxies.
We describe the details of observations and data reduction, and present
an atlas of integrated CO intensity maps, velocity fields and
position-velocity diagrams along the major axes.
 The molecular gas morphology in the Virgo galaxies shows a wealth of
variety, not specifically depending on the Hubble types.
Several galaxies show strong concentration of gas in the central
few kpc region, where the CO morphology shows either "single-peak" or
"twin-peaks".
Morphology of more extended CO components can be classified into
"arm-type",  "bar-type", and  "amorphous-type".

{\bf Key Words} : galaxies: spiral --- galaxies: ISM --- galaxies: structure ---
galaxies: Virgo
\end{abstract}

\section{Introduction}

CO-line observations play an essential role in studying the kinematics and
interstellar physics in the central regions of spiral galaxies,
where the interstellar matter is mostly in the molecular-gas phase and is
strongly concentrated (Sofue et al. 1995; Honma et al. 1995).
There have been numerous observations of nearby galaxies in the CO line
emissions with single dish telescopes (Young \& Scoville 1991;
Braine et al. 1993; Young et al. 1995; Nishiyama \& Nakai 2001).
Large-scale CO line surveys of the Virgo galaxies have been obtained
using the FCRAO 14-m telescope at an angular resolution of 45$''$ by
Kenney \& Young (1988),
and the BTL 7-m telescope by Stark et al. (1986).
These surveys with single dishes were made with angular resolutions of
tens of arcsec, and have given information about the global structure
of molecular disks in Virgo and nearby galaxies.

Interferometer observations at high angular resolutions are
crucial for studying detailed molecular disk structures within the
central few hundred parsecs (Sargent \& Welch 1993).
High-spectral resolution is also crucial to investigate the detailed
kinematics of the central gas disks.
Both high spatial and high spectral resolutions provide us with
precise velocity fields and rotation curves, which are the basis for
deriving the fundamental parameters such as the mass distribution,
bars and related shock phenomena, triggering mechanism of starburst and/or
fueling mechanism of massive black holes.
Interferometer observations have often performed to investigate the
individuality of each galactic center and activity. Recently, some
large surveys of nearby galaxies have started to be reported.

The Nobeyama mm-wave Array (NMA) and Owens Valley Radio Observatory (OVRO)
mm-wave array were used since 1990's to map the central regions of nearby
spiral galaxies with the CO line at a typical angular resolution of
$4\arcsec$ (Sakamoto et al. 1999a).
The Berkely-Illinois-Maryland Association Survey of Nearby Galaxies
(BIMA SONG) has mapped 44 nearby galaxies at typical resolutions
of $6''$ (Regan et al. 2001).
Interferometer observations of several nearby galaxies have been
also conducted from various interests, such as
bars (e.g., Kenney et al. 1992; Regan et al. 1999),
star formation (e.g., Wong \& Blitz 2002), and
nuclear activity (e.g., Baker 1999; Sakamoto et al. 1999a; Kohno et al. 1999;
Schinnerer et al. 1999).

The ViCS (Virgo high-resolution CO survey) project with the NMA has been
performed in order to obtain a homogeneous high angular- and
spectral-resolution database for a large number of CO-bright Virgo
Cluster spirals in the $\co$ line.
Angular resolutions were $\sim 3\arcsec$ after reduction
in the conventional CLEAN procedure with natural weighting.
The major scientific motivation was to investigate the detailed
central kinematics of the galaxies, particularly the innermost rotation
curves from analyses of position-velocity diagrams across the nuclei,
which would be effective to detect central compact massive objects.
The data are also useful for investigation of the kinematics and
ISM physics of the central molecular disks, and their environmental
effect in the cluster circumstance.

An advantage to observe the Virgo Cluster galaxies is their almost identical
distance, which has been accurately determined to be 16.1 Mpc
($1\arcsec$ corresponds to 78 pc) by the Cepheid calibrations
(Ferrarese et al. 1996). Since our target galaxies lie within 2 Mpc
from the Virgo center, M87, the distance ambiguity will be at most 15\%,
mostly less than 10\%.
The accurate distance will enable us to estimate physical quantities
rather precisely, such as the CO and dynamical masses, and linear scales of
gas disks.

The ViCS results will be published in a series of papers. In this paper
we describe the overall observations and reduction, and present an atlas
of the central molecular disks of Virgo galaxies. In the forthcoming
papers we will describe more details of the observations, analyses,
and results for individual galaxies, as well as rotation curves and
central kinematics, investigations of the ISM physics, and comparison
with other wavelengths observations. The database will be opened for
public use on our web page.

\section{Observations and Reduction}

\subsection{Target Galaxies}

The target galaxies in the survey have been selected from the list of
spiral galaxies of the FCRAO CO-line survey by Kenney \& Young (1988)
by the following criteria.

\begin{enumerate}

\item{The sources were chosen in the order of CO line peak antenna
temperatures $T_{\rm A}^{\star}({\rm peak})$ at the optical centers.
Twenty-eight galaxies with the peak antenna temperatures above
$\rm 20\,mK$ were selected from 42 galaxies of Kenney \& Young (1988).}
\item{Inclination angles were limited to be $25\Deg \leq i \leq 75\Deg$
in order to investigate central gas dynamics. This criterion excluded
NGC 4293 ($i=76\Deg$), NGC 4302 ($i=90\Deg$), NGC 4312 ($i=78\Deg$), and
NGC 4710 ($i=90\Deg$).}
\item{Galaxies with morphological type of S0, i.e. NGC 4293, NGC 4710
and NGC 4438, were excluded.}
\item{Interacting galaxies were excluded by a criterion that the
galaxies have no companion within $8'$ radius.
Pairs of NGC 4568/4567, NGC 4298/4302, and NGC 4647 were excluded.}
\item{Peculiar galaxies in optical images, i.e. NGC 4438 and NGC 4064,
were excluded.}
\item{Galaxies observed with the NMA since 1994 were excluded.
NGC 4321 and NGC 4527 have been observed by Sakamoto et al. (1995) and
Sofue et al. (1999), respectively.}
\end{enumerate}

Sixteen galaxies were selected on the basis of these criteria, and we
have observed 15 galaxies except NGC 4450.
The targets are listed in Table 1, which also summarizes the morphological
type, B-band total magnitude, optical size, inclination, position angle
from optical isophotal contours, and nuclear activity from optical
spectroscopy (Ho et al. 1997a,b).
The table list also the CO-line peak temperature,
integrated intensity, intensity-weighted mean velocity, and velocity width
from the single dish CO-line observations (Kenney \& Young 1988),

The selection criterion 1 by the peak antenna temperature was applied
because of higher probability of detection in a single channel.
We note that the FCRAO survey with 45$''$ resolution (Kenney \& Young 1988)
shows that all galaxies, except one, are centrally CO peaked.
Their data show that the peak temperature and maximum CO intensity have
approximately a linear correlation, and that the CO scale radius is
an increasing function of maximum CO intensity.
This implies that our selection by peak temperature is approximately
equivalent to a selection by total integrated CO intensity, and hence
by total CO luminosity for their equal distances.
Hence, our target galaxies, and some that were not selected by the reason
that they were already observed by NMA, would represent the most CO
luminous Virgo galaxies. 

\vskip 3mm
\centerline{--- Table 1   ---}
\vskip 3mm

\subsection{Observations}

We have performed aperture synthesis observations of the $\co$ line
emission from the 15 Virgo galaxies listed in Table 1 in the course
of a long-term project during the winter seasons of 2000 (1999 Dec.
-2000 Apr.), 2001 (2000 Dec.-2001 Apr.) and 2002 (2001 Dec.-2002 Apr.).
We made the observations in three available configurations:
AB (long baselines), C (medium) and D (short) configurations.
The visibility data covered projected baselines from about 10 to 351~m.
The NMA consisted of six 10-m antennas, providing a field of view with
a FWHP beam width of $65\arcsec$ at 115~GHz.
Since interferometry observations sample data in a Fourier space,
the range of collected Fourier components, or baselines, determines
the detectable sizes of objects. In our observations, the antenna size
limited the minimum projected baseline length, restricting the largest
detectable size to about $54''$. Thus our data may miss some fluxes
of extended components of the objects.
Table 2 lists the observation periods and array configurations,
observed central frequencies, and the positions of pointing centers
as well as phase-reference centers for individual galaxies. 

\vskip 3mm
\centerline{--- Table 2   ---}
\vskip 3mm

The antenna were equipped with tuner-less SIS receivers,
which had receiver noise temperatures of about 30 K
in double sidebands, and the typical system noise temperatures were
about 400 K in single sideband.
We used a digital spectro-correlator system (Okumura et al. 2000),
which had two spectroscopic modes (bandwidth of 512 and 1024 MHz);
 we used the mode covering 512 MHz
($1331 \kms$) with 256 channels and 2 MHz ($5.2 \kms$) resolutions.

The nearby radio point source 3C 273 was used as the flux and phase
calibrator, which was observed every 20 minutes.
The band pass response across the channels was also calibrated using
the 3C 273 data. The intrinsic flux density of 3C 273 at the observing
frequency was calibrated for each observing run (typically 5 days)
using the planets (Mars etc.).
The flux of 3C 273 during the three years of observing periods was
gradually variable between 9 and 12 Jy.
The uncertainty in the absolute flux scale for each observing run
was $\sim \pm$ 15 \%, which apply to all results presented
in this paper.

\subsection{Reduction}

The raw data were calibrated using UVPROC-II, a first-stage reduction
software (Tsutumi et al. 1997), and were Fourier-transformed using
the NRAO Astronomical Image Processing System (AIPS).
We applied the CLEAN method with natural weighting to obtain
three dimensional data cubes (RA, DEC, $\vlsr$).
The intensity data were averaged in 2 to 6 bins ($10.4 - 31.2 \kms$)
of the original channels in the spectro-correlator, and the channel
increments were set to 2 to 4 ($10.4 - 20.8 \kms$).
The intensity scale at this stage was in Jy per synthesized beam,
which can be converted to brightness temperature in Kelvin.
The resultant synthesized beam sizes range
in $2-5\arcsec$. The typical rms noise scaled for a $10\kms$ channel was
${\rm 20 mJy \, beam^{-1}}$.
Table 3 lists the resultant parameters of the data cubes for
individual galaxies.

\vskip 3mm
\centerline{--- Table 3   ---}
\vskip 3mm

In addition to the above reduction parameter set, we CLEANed the data
with tapered and uniform weighting functions, which provided
low ($\sim5"$) and  high ($\sim1"$) resolution maps, respectively.
We will present those maps in separated papers for discussing
individual galaxies.

We calculated the fractions of the recovered
single-dish flux from our aperture synthesis observations,
which are listed in table 3.
We first corrected the data cube for primary beam attenuation,
and convolved with a Gaussian single-dish beam of FWHM $45\arcsec$
(comparable to that of the FCRAO survey), and took the flux at the
pointing center of the FCRAO observations.
The recovered fluxes were typically 80\%, which recovered almost all
fluxes within the field of view.

However, a few galaxies showed exceptionally low and high recovered fluxes.
The recovered flux of NGC 4535 was greater than the FCRAO flux by about twice.
We made careful analyses of the raw data for several times:
We made UV data for C, D, C+D, and AB+C+D array configurations,
which were obtained in independent observing periods (Table 2).
We, then CLEANed them separately, but obtained about the same flux for
all the configurations.
Also the rms noises of the reduced cubes and maps are comparable to
those for other galaxies observed in the same periods.
Hence, we conclude that the flux calibration was correct.
The flux disagreement could be possible, if the FCRAO flux
was about significantly under-estimated, and ours was about 15\%
over-estimated, both within the measurement errors.
The recovered flux of NGC 4548 was only 16\% of the FCRAO flux.
This may have happened due to larger sizes of the missing components
than the maximum  detectable size of our observations (\S 2.2).
Alternatively, it could be due to very low brightness of the extended
components.
In fact, the FCRAO flux is as weak as 6.7 K \kmps\ with 45$''$ beam.
If the rest 84\% is extended in the 45$''$ beam,
the intensity for our beam ($2''.6 x 2''.0$) would be 5.6 K \kmps.
For an assumed line width of about 20 \kmps in our beam, the expected
brightness is only 30 mK, which is much below our detection limit. 

We have checked for continuum sources in the galaxies by making channel
maps for a wide range of velocity.
The channel maps for individual galaxies are shown in figure A1 in the
Appendix, where the outernost channels can be used to check continuum fluxes.
As figure A1 shows, no significant continuum source has been detected in
any of the observed galaxies by the present sensitivity, which is 
typically 20 mJy beam$^{-1}$ in rms noise as listed in table 3.
Nevertheless, such galaxy as NGC 4579, which contains an AGN, could 
have some continuum emission. 
So, we applied deeper continuum checking for this galaxy.
We CLEANed the 2 MHz $\times$ 256-channel data cube of NGC 4579
by binning every 32 channels (168 km/s).
However, no continuum source stronger than 10 mJy was found in the 
outermost-velocity channels ($\pm \sim 500 \kms$), where no CO line 
emission is expected. 

\section{The CO Atlas of Virgo Spirals}

\subsection{CO Intensity Maps}
 
Figure 1 (top-left panels) shows optical `looks' of the observed
galaxies for a $5'\times5'$ area taken from the STScI Digitized Sky
Survey (DSS) second generation blue images.
The inserted squares show areas for the CO maps.
We present total integrated intensity maps of the CO emission in the bottom
left panel.
The intensity maps were obtained by using the AIPS task 'MOMNT', which
integrated the intensities by velocity only when the intensity exceeds a
threshold level.
The threshold level was taken to be 2 to 3 times the rms noise in the
data cube.
Channel maps of the observed galaxies are shown in Appendix, and will
be discussed in more detail in the forthcoming papers of this series
on individual galaxies.

The bottom-right panel shows intensity-weighted velocity fields
for the $1'\times 1'$ regions (our field of view is $65\arcsec$ at FWHP
at 115~GHz), and the top-right panels show position-velocity diagrams
along the major axes (top right), except for
NGC 4254 and NGC 4402, for which $80''\times80''$ regions are shown.

The primary-beam correction has not been applied in these maps.
The intensity scales in the maps are in Kelvin of brightness temperature,
rather than in $\rm Jy\,beam^{-1}$ scale that is
directly derived from interferometer observations, for convenience
to compare with single dish observations and to convert to the molecular
hydrogen column density.
We measured the peak CO intensities using these maps and listed in table 5
together with the CO peak brightness temperatures as read from the data cubes.

\vskip 3mm
\centerline{---  Table 5  ---}
\vskip 3mm

Figure 2 shows the CO intensity maps in the same angular and linear scales.
Each box covers a $1' \times 1'$ region, which corresponds to
4.7 kpc~$\times$~4.7~kpc region for an assumed Virgo distance of 16.1 Mpc
(Ferrarese et al. 1996).
Figure 3 shows velocity fields corresponding to figure 2.
In figure 4 we plot the CO intensity maps in the same angular scale
on the Virgo Cluster region, where each map scale has been enlarged by
50 times the real angular size.

\vskip 3mm
\centerline{--- Fig. 1   ---} 
\vskip 3mm
\centerline{--- Fig. 2   ---} 
\vskip 3mm
\centerline{--- Fig. 3   ---} 
\vskip 3mm
\centerline{--- Fig. 4  ---}
\vskip 3mm

\subsection{Velocity Fields}

Figure 1 (bottom left panels) shows intensity-weighted velocity fields
for the observed $1'\times 1'$ regions, which are the same regions for
the integrated-intensity maps at the bottom right panels.
For NGC 4254 and NGC 4402, the $80''\times80''$ regions are presented.
Figure 3 shows the velocity fields in the same angular scales.
The general pattern of the velocity field in figures 1 and 3 is
a symmetric spider diagram, indicating a regular circular rotation
of the CO disk. Slight non-circular streaming motions, such as due
to spiral arms and bars, are superposed on the regular rotation.
However, there are some galaxies that show strong non-circular motion;
NGC 4569 has an extremely large deviation from the circular rotation,
indicating either high-velocity streaming or a large-amplitude warping
in the central disk.

\subsection{Position-Velocity (PV) diagrams}

Position-velocity (PV) diagrams are shown in figure 1 (top right).
These diagrams were made by slicing the data cubes along the optical
major axes with appropriate widths for individual galaxies.
Most galaxies have steeply rising rotation velocity in the central 100
to 200 pc radii.
Such sudden rise of rotation velocity in the close vicinity of the nuclei
had not been clearly detected in the lower-resolution observations.
One of the major purposes of the present CO survey was to obtain the
central rotation curves to investigate possible central massive cores,
which have been found in many nearby galaxies (Sofue et al. 1999;
Takamiya \& Sofue 2000; Sofue and Rubin 2001; Sofue et al 2001;
Koda et al 2002).
In a separated paper (Sofue et al. 2003a), we describe the
result of detailed analyses of the PV diagrams and derivation of accurate
rotation curves by applying a new iteration method (Takamiya \& Sofue 2002),
and discuss the central mass distribution.

\subsection{Uni-scale Atlas of CO Intensities}

In order to give an overview on the general characteristics of
distributions of the molecular gas (CO intensity) in the observed galaxies,
it is helpful to compare the galaxies in a unified scheme.
In figure 2, we present the observed $I_{\rm CO}$ distributions in the same
angular and intensity scales.
The image sizes are all $1.'0 \times 1.'0$, corresponding to
4.68~kpc $\times$ 4.68~kpc for an assumed distance of 16.1 Mpc.
The contours are drawn at the same levels of 5, 10, 20, 40, 80,
160 $\rm K \, \kms$ for all galaxies.

\subsection{Uni-scale Atlas of Velocity Fields}

Figure 3 shows the same as Figure 2, but for distributions of intensity
weighted velocities.
The velocity fields generally show a 'spider diagram' pattern,
indicating circular rotation of the molecular disk.
The rotation velocity rises rapidly in the central few hundred parsecs,
which is more clearly observed in the position-velocity diagrams in
figure 1.
In many galaxies, the spider diagrams are more or less distorted,
indicating either non-circular streaming motion or warping of the gas
disk.

\subsection{Sky Plot of CO Maps on the Virgo Cluster region}

In figure 4 we plot the $I_{\rm CO}$ maps on the sky area of the Virgo
Cluster in a similar manner to a plot of HI maps by Cayatte et al. (1990).
The angular scales are enlarged by a factor of 50 for individual galaxies.
The CO distributions appear to be not strongly correlated with the distance
from the center of the Cluster at M87, which is marked by a cross.
This property is very different from the HI gas distribution;
The HI disks of inner-cluster galaxies are usually largely distorted and
are often truncated by the ram-pressure of the intracluster
medium (Cayatte et al. 1990), while the central CO disks are not
strongly perturbed by the ambient gas in the cluster, probably because
they lie deep in their galactic potentials.

\subsection{The ViCS Data Base}

The calibrated and reduced data presented in this paper will be opened
on our web page in the form of FITS formatted cubes and maps, and in
gif-formatted images at the URL,
http://www.ioa.s.u-tokyo.ac.jp/radio/virgo/.

\section{Central Positions and CO Distributions}

\subsection{Central Positions}\label{sec:dycen}

In order to determine the central positions, we fitted a disk model
with a Brandt-type rotation curve (Brandt 1965) to
intensity-weighted velocity fields using the AIPS task GAL.
Since this task assumes a pure circular rotation, we used only
central several arcseconds for the fitting, out of which
the isovelocity contours indicate some deviations from circular rotation.
The iteration in the task could occasionally provide different results
with different initial guesses.
We checked this error by changing the initial guess in appropriate ranges,
and confirmed that the task suffices to provide the dynamical centers
accurate to about $1\arcsec$ in most cases.
The derived central positions are listed in Table 4. 

The thus obtained center positions were in coincident with the
NED center positions within an arcsecond in most cases.
However, in such cases that the central CO distribution is not smooth,
or the velocity field is strongly perturbed, and hence above dynamical
centers are not reliable, we determined the central positions from
the literature.
The center of NGC 4212, which shows patchy CO distribution (Figure 1),
was adopted from the optical observations by Cotton et al. (1999),
which coincides with our CO emission peak within an error of
$\sim1\arcsec$. NGC 4569 shows strong noncircular motions in the
velocity field and position-velocity diagram; we adopted the central
position determined by Sakamoto et el. (1999) from their CO
interferometry observations with a lower resolution.
The center of NGC 4579 was taken to coincide with the position of the
unresolved radio continuum source (Ho and Ulvestad 2001).

\vskip 3mm
\centerline{---  Table 4  ---}
\vskip 3mm

\subsection{Radial Distribution of CO gas}

Figure 5 displays azimuthally averaged radial profiles of
CO-line intensities in unit of K $\kms$ as projected on the galaxies' disks
corrected for the inclinations.
In order to make these plots, integrated-intensity maps
without clipping were corrected for the primary beam attenuation,
and we applied the AIPS task IRING around the central positions derived
in \S \ref{sec:dycen}.
We fixed the inclination and position angles for each galaxy to those
provided from optical observations (Table 1).

The sampling intervals in the plots were set to be $0''.5$.
However, the effective sampling intervals are equal to the beam widths.
The number of effective sampling points for the fit increases with radius $r$
proportionally to $r$, and hence the statistical error decreases
proportionally to $r^{-1/2}$.
The intensity error at each point in a map is given by
$\Delta I = \Delta T \times N_{\rm c}^{1/2}\times \Delta V$
($\sim 15$ K \kmps) ,
where $\Delta T (\sim 0.3$ K), $N_{\rm c} (\sim 25)$,
and $\Delta V (\sim 10 \kms)$ are the rms,  number of channels
within the expected velocity width, and   the velocity
interval, respectively, as given in  table 3.
Therefore, the typical error of the profiles is given approximately by
$ \sim 15(r/{\rm beam~width})^{-1/2}$ K \kmps.

\vskip 3mm
\centerline{---  Fig. 5  ---}
\vskip 3mm

The intensity distributions in the central 10 to 15$''$ regions are
approximately exponential with scale radii 5 to 10$''$, or 400 to 800 pc.
These scale radii are a few times smaller than those derived by Regan
et al. (2001). Since our survey has three times higher resolution (2")
than theirs (6"), we may detect the central cusps of the CO distributions.
In most cases, the outer regions than 20$''$ are not significantly detected,
except for some cases with disk components.
Three of our target galaxies, NGC 4254, 4501 and 4569, have been
observed in CO line with lower sensitivity and resolutions (Sakamoto
et al. 1999). Our radial profiles for NGC 4501 and 4569 are consistent
with those from the previous observations.
The profile for NGC 4254 is also roughly consistent
with the previous result, while there are some small scale deviations. 

The ellipse fit gives a quantitative presentation of the radial profiles
including the outskirts.
However, the fit looses linear resolution, depending strongly
on the inclination, because it uses the data in the minor axis direction
with an equal weight.
Figure 5 shows that many galaxies have a strong concentration of CO gas
within the central 10$''$ (0.8 kpc) radius, while
some have a plateau or a dip at the center.
We discuss such galaxies in \S 5
as a central-/single-peak and twin-peaks types, respectively.

\section{Molecular gas morphology}

The molecular gas distributions show a wealth of variety.
Although it is difficult to categorize them in a simple way,
we can find some characteristic types in the central gas distributions.
Many galaxies have high concentration of CO gas in the central
a few kpc region, where the CO morphology shows either single peak
or twin peaks.
More extended components have more variety of morphologies,
which can be classified into arm type, bar type, and amorphous type.

\subsection{Central gas distributions}

\begin{enumerate}
\item{Central peak and/or Single peak:}
Many galaxies show strong concentration of molecular gas around the nuclei
in so far as the present maps are concerned.
NGC 4212, NGC 4419, NGC 4501, NGC 4535, and NGC 4536 are the examples.
The typical size of these central peaks is about 200 - 400 pc.
In most cases, their peaks are single at the present resolution,
which we call "single peak".
Figure 6 shows the CO intensity distributions in the central $20'' \times 20''$
regions (1.6 kpc square) of the central-/single-peak galaxies.
The central/single peak galaxies shares a considerable fraction
among the observed galaxies.
Note that the object selection was made by peak antenna
temperature in the FCRAO survey with a 45$''$ beam.
Hence, the present maps could have a selection effect for galaxies with
higher peak-temperatures.
However, as argued in \S 2, our object selection is approximately equivalent
to a selection by CO luminosity.
We may consider that the statistics
with the presently observed galaxies is significant to discuss the general
types of central CO morphology of the most CO luminous Virgo galaxies.

\vskip 3mm
\centerline{--- Fig. 6 ---}
\vskip 3mm

\item{Twin peaks:}
Typical example of twin-peak molecular gas distribution is seen for
NGC 4303, which shows two offset open molecular arms along the optical
bar, which end at a molecular ring with two peaks.
Kenney et al. (1992) have reported both single-peak, and twin-peaks types.
They selected four barred galaxies with strong CO and FIR emission,
and showed that the barred galaxies have twin peaks in CO likely
as the consequence of bar-induced inflow.
In so far as the present data set is concerned, which includes galaxies of
random types with the CO emission concentrated in the central
$45''$ (3.5 kpc) regions, twin-peak galaxies shares rather a small fraction.
\end{enumerate}

Kenney et al. (1992) showed that the separations between twin peaks
are about 200 - 400 pc, while our single peak galaxies do not have
double peaks even in the same scale ($3''\sim200\,{\rm pc}$).
Note, however, that this classification may depend on the spatial
resolution; it may happen that a single peak at our resolution
consists of a more number of inner structures at higher resolution.
In fact, NGC 4501 appears to be a single peak type in the present atlas,
while a higher resolution image show a small patchy ring with a diameter
of $\sim 3"$ (Onodera et al. 2003:private communication).

PV diagrams may apparently imply spatially-unresolved double peaks with
separated velocities. However, a single peak galaxy may apparently show two
peaks on a PV diagram, at the positive and negative terminal velocity ends
at turnover radii of central rising and outer flat rotation curves,
even when the gas distribution has no spatially-separated double peaks
(Sofue et al. 1999; Sakamoto et al. 1999a).
Hence, PV diagrams may not be used for the spatial morphological
classification.

\section{Description of Individual Galaxies}

We describe individual galaxies about their CO properties obtained from
the present observations.

\subsection{ NGC 4192}
An extremely bright CO peak is observed at the center, which classifies this
galaxy in a "central-peak" type, while it appears that the peak may be
resolved into two peaks at higher resolution. Hence, the classification depends
on the resolution. The central peak is surrounded by a bright disk at high
inclination at about the same inclination as the optical disk.
The position-velocity diagram shows  a very high-velocity rotation,
whose maximum reaches almost $250 \kms$.
However, the central CO peak is rotating more slowly at about $100 \kms$.

\subsection{ NGC 4212}
The CO distribution consists of a molecular core and extended straight
arms in the direction of the major axis.
The core is shifted from the map center toward the NE by a few arcseconds.
The velocity field and PV diagram indicate that the rotation is rigid
body-like, while the core shows  a steeper velocity gradient.

\subsection{ NGC 4254}
The central molecular gas distribution shows a bar-like elongation,
while no optical bar feature is seen in the visual-band images.
The CO intensity has a slight depression at the dynamical center,
which coincides with the nucleus.
Two well-developed spiral arms wind out from the bar
ends toward the south and north.
The south-eastern arm bifurcates into a tightly-wound dense molecular
arm with an almost zero pitch angle.
Hence, the molecular disk has three arms, and the
arms are well correlated  with optical dark lanes.
The velocity field shows a regular spider pattern, indicating a circular
rotation of the disk, on which small-amplitude streaming motion due to the
spiral arms are superposed.
The PV diagram shows a sharp rise in the central few arcseconds,
indicating a massive core, and then the velocity increases gradually.
Overall distributions and kinematics agree with the previous low resolution
observations (Sakamoto et al.\ 1999a).
A detailed study of this galaxy with consideration of the ram-pressure
effect by the intra-cluster medium is presented in Sofue et al. (2003b).

\subsection{ NGC 4303}

CO gas is highly concentrated in the nuclear disk within a radius
$r \sim 8''$ (600 pc).
The nuclear disk comprises the "twin peaks" at the eastern and
western edges of the nuclear disk, and there appears to exist a
diffuse central component around the nucleus between the twin peaks.
Two prominent bisymmetric spiral arms, or offset ridges, wind out from these
twin peaks, and extend toward the north and south along the dark lanes in the
optical bar. 
The PV diagram along the major axis (north-south) indicates a rise
of rotation velocity within $r \sim 2''$ (160 pc) to 160 - 180 \kmps.
Our result is consistent with the high-resolution observations with the
OVRO interferometer by Schinnerer et al. (2002).
Detailed description of this galaxy are given in a separate paper of this
series by Koda et al. (2003: private communication).

\subsection{NGC 4402}
The CO intensity distribution shows a high density nuclear molecular
disk of $r \sim 10''$.
The nuclear disk is surrounded by a more extended molecular disk of radius
$\sim 30''$ (2 kpc).
This outer disk appears to be consisting of two spiral arms,
one extends to the west from the southern edge of the nuclear disk,
tracing the dark lane, and the other arm toward the east from the
north-eastern edge of the nuclear disk. 
The velocity field shows a usual spider diagram superposed by some
streaming motion in the molecular ring/arms.
The PV diagram shows a nuclear component and outer ring/arms.

\subsection{ NGC 4419}
The CO gas is strongly concentrated in the central $5''$ radius disk,
which is associated with an elongated outer disk component.
Kenney et al. (1990) also reported the concentrated CO distribution.
The outer molecular disk is lopsided toward the north-west.
The PV diagram indicates that the central component has a rotation
velocity as high as 150 \kmps within 5$''$ radius, which is followed by a
gradually rising disk rotation.

\subsection{ NGC 4501}
A nuclear concentration of the molecular gas of radius $3''$ is remarkable,
as reported by Sakamoto et al.\ (1999a).
This is classified as the "single-peak" type, although the peak intensity
is not particularly high compared with the other typical single peaks.
This central peak is surrounded by an extended component elongated
in the SE to NW direction, with the SE end at 5$''$ radius being brighter.
Two prominent molecular arms are running at $r \sim 20''$.
The north-eastern arm is much stronger than the south-western arm.
Both arms are associated with the dark lanes along the optical spiral arms.
The velocity field and PV diagram indicate sharp rise of rotation
velocity in the nuclear disk.
Detailed description and modeling by spiral-shock accretion mechanism
are given in Onodera et al. (2003:private communication).

\subsection{ NGC 4535}
The molecular gas shows a strong concentration in the central region
of $\sim 6''$ radius. This galaxy is a typical "single-peak" type.
Offset bars are extending from the central disk toward the
NE and SW, coinciding with the optical dark lanes in the bar.
The velocity field shows a usual spider diagram
with the zero velocity node at position
angle of $90\deg$, coinciding with the optical minor axis.
However, the CO arms along the dark lanes show some non-circular
streaming velocity, indicating inflow along the arms.
The PV diagram shows a sharply rising, but rigid body-like behavior
within the central molecular disk.

\subsection{ NGC 4536}
This is also a typical "single-peak" type galaxy with the molecular gas
being  concentrated in the nuclear disk of $\sim 10''$ radius and an
unresolved compact core exists at the nucleus.
The velocity field shows a  spider diagram, and the PV diagram
indicates that the rotation velocity rises to 200 \kmps
within the central $2''$.
There appears no strong non-circular streaming motion.

\subsection{ NGC 4548} 

This is a typical barred galaxy.
CO emission is very weak compared with the other galaxies.
The CO distribution is highly concentrated near the nucleus, being
centrally peaked, and no extended emission is detected.
The recovered flux is only 16\% of the FCRAO flux with 45\" beam
(\S 2.3), and the rest 84\% (5.6 K \kmps) could be due to very
extended components with sizes greater than our maximum
detectable size (54$''$). 

\subsection{ NGC 4569}
 
The molecular gas is highly concentrated within $\sim 1$ kpc radius.
The CO intensity distribution is elongated in the same direction as the
optical major axis, and has two peaks with depression at the nucleus,
consistent with the earlier CO map (Sakamoto et al.\ 1999a).
Thus, the central molecular morphology of this galaxy may be classified
in twin peaks at the present resolution.
However, a higher resolution CO map reveals that the apparent
two peaks coincide with both ends of an elliptical molecular ring,
while they are not associated with so called offset ridges
(Nakanishi et al. 2003).
The velocity field is strongly disturbed from circular rotation, and
the PV diagram indicates significant `forbidden' velocities.
Nakanishi et al.(2003) discuss the kinematics of this galaxy in
detail, and tried to explain these features using two models of non-circular
motion and warping of the inner disk. And they conclude that it is
natural that disturbed velocity field and forbidden velocities of the
PV diagram are due to non-circular motion.
Helfer et al.\ (2001) have reported an extended CO emission
from their wide-field mosaic image. Jogee et al.\ (2001)
have also presented a high resolution CO image of this galaxy. 

\subsection{ NGC 4571}
No significant detection of the CO line was obtained for this galaxy,
not only in channel maps, but also in an integrated-intensity map.
Our resultant rms was $11.5\,K\,\kms$ for a $130 \kms$ width, that is
greater than $3.3\,K\,\kms$ from single dish observations (Table 1).
Since the map shows only noises, we do not present the result.

\subsection{ NGC 4579}
The CO distribution is elongated in the east-west direction, about $30\deg$
displaced from the optical bar axis.
There are two major CO peaks with asymmetric peak intensities, which are
associated with symmetric spiral features
as reported by Kohno et al.\ (1999).
The velocity field shows a higher rotation velocity than 200 \kmps
in the central few arcseconds, which is more clearly visible in the
position-velocity diagram.

\subsection{ NGC 4654}
This galaxy is known for its lopsided structure in the optical as
well HI disks, most likely due to the ram-pressure effect of the
intracluster gas, blowing from the northwest (Phookun and Mundy 1995).
The CO distribution is also lopsided,
in the same direction as that of the HI and optical image tail.
The lopsided CO distribution suggests that the ram pressure
effect is not negligible even in such a central molecular disk.
Moreover, the CO distribution is more elongated than the optical/HI disks.
The velocity structure is rigid-body like with mildly increasing rotation
velocity with the radius.
Such rotation characteristics is exceptional among the presently
observed PVDs.

\subsection{ NGC 4689}
Like the optical spiral arm features, the CO intensity distribution is
amorphous, and is patchy and widely extended.
There appears neither spiral arms nor bars in CO, and no central peaks are
found.
The peak $I_{\rm CO}$ amounts only to $\sim 24 {\rm K \kms}$,
the lowest among the observed galaxies.
The velocity field indicates a regular rotation pattern, but the central
rise of rotation velocity is mild, as indicated by the PV diagram.

\section{Summary and Discussion}

We have obtained high-resolution CO-line survey of 15 Virgo spiral galaxies
using the Nobeyama Millimeter-wave Array in AB, C and D configurations,
and presented the result in the forms of integrated intensities, velocity
fields and position-velocity diagrams along the major axes.
The galaxies were sampled from the CO brightest galaxies in the Kenney and
Young's (1988) list without any bias.
The CO properties may be compared with each other without ambiguity of
the linear scale, as the distance to the Virgo cluster galaxies are safely
taken to be 16.1 Mpc from the Cepheid calibration (Ferrarese et al. 1996).
For the homogeneity, our data will be useful for investigating correlation
between the CO properties and other characteristics. 
We will discuss the correlation of the central peaked CO distributions
and nuclear activities in a forthcoming paper of this series.
In the second paper of this series, we will derive exact rotation curves
by analyzing the position-velocity diagrams, and discuss the
dynamical properties of the central regions of the Virgo galaxies,
as well as detailed mass distributions.

We summarize the results obtained in this paper as follows,
and discuss the implications below.
The mean radial profiles of the molecular gas distribution in the inner
10 to 15$''$ radius regions are approximately exponential with $e$-folding
scale radius of 400 to 700 pc.
In some galaxies, more extended disk components are detected, whose
scale radii are greater, while the present interferometry
data are not appropriate to determine the disk radii precisely.

A careful inspection of the intensity maps shows that the observed
intensity distributions have a variety of types,
which  may be classified into the following types.
The centrally concentrated components can be classified into two types:
the central-peak or single-peak type and  twin peaks type.
The latter shows plateau-like radial profiles near the center.
The distributions of more extended components can be classified into
spiral arms, bars, and amorphous types.
It is of particular interest to consider what causes the variety
of molecular gas morphology, as it could be intimately related
to the activities in the centers of galaxies.

Twin-peaks of molecular gas at the ends of a set of two bisymmetric
offset molecular bars (dark lanes) along the optical bar have been
noticed for several barred galaxies in relation to fueling mechanisms
of interstellar gas toward the central regions (Kenney et al. 1992;
Sakamoto et al. 1999a).
NGC 4303 is a typical case for the twin-peaks type, and the structure is
well explained by a bar-potential and galactic shock hypothesis (Schinnerer
et al 2002; Koda et al 2003:private communication).
However, the fraction of galaxies having "twin-peaks" is not particularly
high in so far as the present resolution maps ($2-4''$ or
150 - 300 pc) are concerned.

On the other hand, "central-" or "single-peak" galaxies share
a larger fraction: five galaxies among the fifteen in the present survey
at our resolution.
Examples are NGC 4192 (SAB(s)ab)), NGC 4419 (SB(s)a), NGC 4501 (SA(rs)b),
NGC 4535 (SAB(s)c) and NGC 4536 (SAB(rs)bc).

Twin-peaks are often thought to be a consequence of characteristic gaseous
orbits in a bar potential: x1 and x2-orbits in a bar intersect each other
at the inner ILR, causing collision of the gas at the two intersecting
points, and consequently producing twin peaks. Although this simple
interpretation of the twin-peaks is attractive, the current study has shown
that the single-peak is more popular than the twin-peaks.
It is an interesting subject to consider the origin of the single-peak,
while we need a more careful simulation and gas dynamics.
We here try to speculate a possible mechanism.

First of all, the twin-peaks are not the final and stable structure of
the gas distribution in a bar.
The selfgravity of the gas will cause a collapse in the gas structures,
particularly such large clumps as twin-peaks will be gravitationally
unstable, and cause further infall of gas into the center due to the
friction among the clouds (Wada \& Habe 1992).
Moreover, if there exists central massive object, the friction due to
stronger shearing motion will accelerate the accretion toward the
nucleus (Fukuda et al. 1998).
In fact, most of the galaxies show very steep rise of rotation curve
in the central $\sim 100$ pc region, indicating the existence of
massive compact cores of mass $10^8 -10^9 M_\odot$ around the nuclei
(Sofue et al. 2003a).
Hence the central single-peak may be formed after the twin-peaks
have developed.

The above mechanism could work even if a galaxy shows no prominent bar in
the optical/infrared photographs, because even a very weak bisymmetric
distortion of a disk potential can cause non-circular motion of gas
(Koda \& Wada 2003). Also, Onodera et al. (2003: private communication)
discuss another possible mechanism to produce a central single-peak by
stellar spiral arms for the case of NGC 4501, which indeed has no bar,
but shows continuous spiral structure from the disk to the nucleus.

\vskip 10mm

Acknowledgements:
The observations were performed as a long-term project from 1999 December
till 2002 April at the Nobeyama Radio Observatory (NRO) of the National
Astronomical Observatories of Japan. We are indebted to the staff of NRO
for their help during the observations. We thank  T. Takamiya and M. Hidaka
for their help with the observations and reductions,  A. Kawamura and
M. Honma for their help with the observations. The data reductions were
made using the NRAO AIPS package. We made use of the data archive from
NASA Extragalactic Database (NED). J. K. was financially supported by the
Japan Society for the Promotion of Science (JSPS) for Young Scientists.

\newpage


\noindent{\bf References}
\vskip 3mm

\bibitem Baker, A. J. 1999,
in The Physics and Chemistry of the Interstellar Medium,
ed.\ V.~OssenKopf, J.~Stutzki, \& G.~Winnewisser (GCA-Verlag, Herdecke), 30

\bibitem Braine, J., Combes, F., Casoli, F., Dupraz, C., G\'elin, M.,
Klein, M., Wielebinsky, R., \& Brouillet, N.  1993, A\&AS, 97, 1791

\bibitem Cayatte, V., van Gorkom, J. H., Balkowski, C., and Kotanye, C.
1990 AJ 100, 604.

\bibitem Cotton, W. D., Condon, J. J., \& Arbizzani, E.\ 1999, ApJS, 125, 409

\bibitem Ferrarese, L., Freedman, W. L., Hill, R. J., Saha, A.,  Madore,
B. F.,
 et al. ApJ. 1996, 464, 568

\bibitem Fukuda, H., Wada, K.,  Habe, A. 1998, MNRAS 295, 463.

\bibitem Helfer, T., Regan, M. W., Thornley, M., Wong, T.,
Sheth, K., Vogel, S. N., Bock, D. C. J., Blitz, L., and Harris, A.
2001, ApSS, 276, 1131

\bibitem Ho, L. C., Filippenko, A. V., \& Sargent W. L. W. 1997a,
ApJS 112, 315

\bibitem Ho, L. C., Filippenko, A. V., \& Sargent W. L. W. 1997b,
ApJ 487, 591

\bibitem Ho, L. C., \& Ulvestad, J. S.\ 2001, ApJS, 133, 77

\bibitem Honma, M., Sofue, Y., \& Arimoto, N., AAp 304, 1

\bibitem Hummel, E., Davies, R. D., Pedlar, A., Wolstencroft, R. D.,
van der Hulst, J. M  1988 A\&A 199, 91

\bibitem Jogee, S., Baker, A. J., Sakamoto, K., Scoville, N. Z., and
Kenney, J. D. P.  2001, in The Central Kiloparsec of Starbursts and AGN:
The La Palma Connection, ASP Conf. Ser. Vol. 249, ed. J. H. Knapen,
J. E. Beckman, I. Shlosman, and T. J. Mahoney
(ASP, San Fransisco), 612

\bibitem Kenney, J. D., \& Young, J. S., 1988, ApJS 66, 261

\bibitem Kenney, J. D. P., Young, J. S., Hasegawa, T., Nakai, N.  1990,
ApJ, 353, 460

\bibitem Kenney, J. D. P., Wilson, C. D., Scoville, N. Z., Devereux, N. A.,
\& Young, J. S. 1992, ApJ, 395, L79

\bibitem Koda, K.,  Sofue, Y., Kohno, K., Nakanishi, H., Onodera, S.,
Okumura, S.K. and Irwin, Judith A. 2002 ApJ. 573, 105.


\bibitem Koda, J., Wada, K. 2003 AA, in press.

\bibitem Kohno, K., Kawabe, R., \& Vila-Vilar\'o, B. 1999,
in The Physics and Chemistry of the Interstellar Medium,
ed. V. OssenKopf, J. Stutzki, \& G. Winnewisser (GCA-Verlag, Herdecke), p.34

\bibitem Kohno, K., Vila-Vilaro, B., Kawabe, R., Sakamoto, S.,
Ishizuki, S., and Matsushita, S.  2002, PASJ, submitted

\bibitem Nakanishi, H., Sofue, Y., Koda, J., and Onodera, S.
2003 PASJ submitted.

\bibitem Nishiyama, K., Nakai, N.,  2001 PASJ  53, 713

\bibitem Nishiyama, K., Nakai, N., \& Kuno, N.  2001 PASJ  53, 757

\bibitem Okumura, S. K.,  Momose, M., Kawaguchi, N. et al. 2000 PASJ 52, 339


\bibitem Phookun, B., Mundy, L. G. 1995 ApJ 453, 154.

\bibitem  Regan, M W., Thornley, M D., Helfer, T T., et al. 2001 ApJ
561, 218

\bibitem  Saikia, D. J., Junor, W., Cornwell, T. J., Muxlow, T. W. B.,
Shastri, P. 1990 MNRAS 245, 408

\bibitem Sakamoto, K., Okumura, S. K., Ishizuki, S., \& Scoville, N. Z.,
1999a, ApJS 124, 403

\bibitem Sakamoto, K., Okumura, S. K., Ishizuki, S., \& Scoville, N. Z.,
1999b, ApJ, 525, 691

\bibitem Schinnerer, E., Eckart, A., \& Tacconi, L. J.  1999,
ApJ, 524, L5

\bibitem Schinnerer, E.,  Maciejewski, W.,  Scoville, N.Z., Moustakas, L.A.
2002, ApJ, in press

\bibitem Sargent, A. I. Welch, W. J. 1993 ARA\&A 31, 297

\bibitem Sofue, Y., Honma, M., \& Arimoto, N., 1995, AAp 296, 33.

\bibitem Sofue, Y., Tomita, A., Honma, M., \& Tutui, Y. 1999, PASJ, 51,
 737

\bibitem Sofue, Y., Koda, J., Nakanishi, H., Onodera, S. M.
2003a PASJ submitted.

\bibitem Sofue, Y., Koda, J., Nakanishi, H., Onodera, S., Hidaka, M.
2003b PASJ in this volume.

\bibitem Sofue, Y., Koda, J., Kohno, K., Okumura, S. K., Honma, M.,
Kawamura, A. and Irwin, J. A. 2001 ApJ 547  L115.

\bibitem Sofue, Y. and Rubin, V.
2001, Ann. Rev. Astron. Astrophys. 39, 137.

\bibitem Sofue, Y., Tutui, Y., Honma, M., Tomita, A.,  Takamiya, T.,
Koda, J., and Takeda, Y. 1999 ApJ.  523, 136.

\bibitem Stark, A. A, Knapp, G. R., Bally, J., Wilson, R.  W., Penzias, A. A.,
Rowe, H. 1986 ApJ 310, 660

\bibitem Takamiya, T, \& Sofue, Y., 2000, ApJ 534, 670

\bibitem Takamiya, T, \& Sofue, Y., 2002, ApJ Letters, submitted

\bibitem Terashima, Y., Iyomoto, N., Ho, L. C., \& Ptak, A. F.  2002,
ApJS, 139, 1

\bibitem Wada, K., \& Habe, A. 1992, MNRAS, 258, 82

\bibitem Wong, T., \& Blitz, L.  2002, ApJ, 569, 157

\bibitem Young, J. S. Scoville, N. Z. 1991 ARA\&A 29, 581

\bibitem  Young, J. S., Xie, S., Tacconi, L., et al. 1995 ApJS, 98, 219


\newpage


\begin{table*}
\small{
\caption{Object List for the High-Resolution
Virgo CO Survey (ViCS)}\label{tab:objlst}
  \begin{center}
    \begin{tabular}{ccccccccccccc}
\hline\hline
NGC   &   Morphology & Activity & $B_{\rm total}$ & Opt. Size   & $i$ & PA & $T_{\rm MB,peak}$  & $I_{\rm CO}$  & $V_{\rm CO}$ & $\Delta V_{\rm CO}$\\
 &   & &  [mag] & [$\arcmin \times \arcmin$] & [$\arcdeg$] & [$\arcdeg$]& [mK]  &  [$\rm K~ km~s^{-1}$] & [$\rm km~s^{-1}$] & [$\rm km~s^{-1}$] \\
(1) & (2) & (3) & (4) & (5) & (6) & (7) & (8) & (9) & (10) & (11) \\
\hline
4192 & SAB(s)ab          & T2                 & 10.95 & 9.8$\times$2.8 & 74 & 155 & 38 &  8.9 & $-$105 & 330 & \\
4212 & SAc               & H                  & 11.83 & 3.2$\times$1.9 & 47 &  75 & 53 &  7.5 &  $-$60 & 130 & \\
4254 & SA(s)c            & H                  & 10.44 & 5.4$\times$4.7 & 28 &  45 & 162& 19.3 &   2407 & 200& \\
4303 & SAB(rs)bc         & H                  & 10.18 & 6.5$\times$5.8 & 25 &   0 & 167& 21.6 &   1555 & 160 & \\
4402 & Sb                & --                 & 12.55 & 3.9$\times$1.1 & 75 &  90 & 96 & 12.4 &    244 & 200 & \\
\\
4419 & SB(s)a            & T2                 & 12.08 & 3.3$\times$1.1 & 67 & 133 & 80 & 18.5 & $-$203 & 300 & \\
4501 & SA(rs)b           & S2                 & 10.36 & 6.9$\times$3.7 & 58 & 140 & 73 & 21.8 &   2263 & 450 & \\
4535 & SAB(s)c           & H                  & 10.59 & 7.1$\times$5.0 & 43 &   0 & 60 &  8.5 &   1949 & 250 \\
4536 & SAB(rs)bc         & H                  & 11.16 & 7.6$\times$3.2 & 67 & 116 & 80 & 18.4 &   1793 & 310 & \\
4548 & SBb(rs)           & L2                 & 10.96 & 5.4$\times$4.3 & 37 & 150 & 44 &  6.7 &    450 & 210 & \\
\\
4569 & SAB(rs)ab         & T2                 & 10.26 & 9.5$\times$4.4 & 63 & 23  & 124& 26.5 & $-$203 & 320 & \\
4571 & SA(r)d            & --                 & 11.82 & 3.6$\times$3.2 & 38 & 55  & 42 &  3.3 &    346 & 130 & \\
4579 & SAB(rs)b   &  S1.9/L1.9                & 10.48 & 5.9$\times$4.7 & 37 & 60  & 58 &  7.5 &   1465 & 380 & \\
4654 & SAB(rs)cd         & H                  & 11.10 & 4.9$\times$2.8 & 52 & 125 & 65 &  6.0 &   1075 & 180 & \\
4689 & SA(rs)bc          & H                  & 11.60 & 4.3$\times$3.5 & 30 & 160 & 51 &  6.5 &   1614 & 210 & \\
\hline
    \end{tabular}
  \end{center}
Notes. ---
Col.(1): Galaxy name.
Col.(2): Morphological type from NED.
Col.(3): Nuclear activity from optical spectroscopy (Ho et al.\ 1997a).
H, S, L, and T mean HII, Seyfert, LINER, and Transient-type (HII + LINER),
respectively.
Cols.(4) and (5) : Total B-band magnitude and apparent sizes along major
and minor axes, taken from NED.
Col.(6)(7): Inclination and position angle of the major axes taken
from Kenney \& Young (1988).
Cols.(8) and (9): CO Peak temperature and integrated intensity at optical
centers taken from the FCRAO survey by Kenney \& Young (1988).
Scaled to main-beam temperature by a scaling factor of
$T^*_{\rm A}/T_{\rm MB}=0.55$ for the FCRAO telescope (Young et al.\ 1995).
Cols.(10) and (11): Intensity-weighted mean velocity and full width at 0~\%
intensity of CO line from Kenney \& Young (1988)\\
}
\end{table*}

\clearpage


\begin{table}
  \caption{Observational Parameters}\label{tab:obsparm}
  \begin{center}
    \begin{tabular}{ccccccc}
\hline \hline
    &         &      & \multicolumn{2}{c}{Field Center} &  Frequency &          \\
NGC & Config. & Year & \multicolumn{2}{c}{(4)}          &   (GHz)    & Reference\\
\cline{4-5}
(1)  &  (2)   & (3)  & RA(J2000)    & DEC(J2000)        &   (5)      &   (6)    \\
\hline
4192 & AB+C+D & 2000-2002 & 12 13 48.30 & +14 54 02.9 & 115.350000 &  1         \\
4212 &  C+D   & 2000-2001 & 12 15 39.11 & +13 54 04.8 & 115.286574 &  2         \\
4254 & AB+C+D & 2000      & 12 18 50.03 & +14 24 52.8 & 114.355710 &  3         \\
4303 & AB+C+D & 2000      & 12 21 54.87 & +04 28 24.9 & 114.659250 &  2         \\
4402 & AB+C+D & 2001-2002 & 12 26 07.06 & +13 06 45.7 & 115.200000 &  1         \\
\\
4419 & AB+C+D & 2000-2002 & 12 26 56.43 & +15 02 51.1 & 115.286574 &  1         \\
4501 & AB+C+D & 2001-2002 & 12 31 59.14 & +14 25 12.9 & 114.407000 &  4         \\
4535 & AB+C+D & 2000-2001 & 12 34 20.25 & +08 11 52.2 & 114.507290 &  5         \\
4536 & AB+C+D & 2001      & 12 34 27.07 & +02 11 18.3 & 114.586000 &  5         \\
4548 & AB+C+D & 2001      & 12 35 26.40 & +14 29 47.0 & 115.098000 &  3         \\
\\
4569 & AB+C+D & 2000-2002 & 12 36 49.82 & +13 09 45.8 & 115.286574 &  4         \\
4571 &  C+D   & 2001-2002 & 12 36 56.40 & +14 13 02.0 & 115.138000 &  3         \\
4579 & AB+C+D & 2001-2002 & 12 37 43.53 & +11 49 05.4 & 114.710000 &  6         \\
4654 & AB+C+D & 2000-2002 & 12 43 55.74 & +13 07 44.2 & 114.811630 &  1         \\
4689 &  C+D   & 2000-2002 & 12 47 45.60 & +13 45 46.0 & 114.659250 &  3         \\
\hline
    \end{tabular}
  \end{center}
Notes. ---
Col.(1): Galaxy name.
Col.(2): NMA configulations used for the observations.
Col.(3): Year of the observations.
Col.(4): Pointing position as well as the phase tracking center.
Col.(5): Observing frequency.
Col.(6): References for positions:
(1) Condon et al. 1990;
(2) Saikia et al. 1994;
(3) NED;
(4) Sakamoto et al. 1999;
(5) Hummel et al. 1987;
(6) Kohno et al. 1999
\end{table}

\clearpage


\begin{table}
  \caption{Parameters of Cubes}\label{tab:obsparm}
  \begin{center}
    \begin{tabular}{ccccccccccccc}
\hline \hline
& \multicolumn{3}{c}{Synth. Beam} & & \multicolumn{2}{c}{$\Delta V$} & & \multicolumn{2}{c}{{\small Rms for $\rm 10\,km\,s^{-1}$}} &  {\small $T_b$ for ${\rm 1Jy\,beam^{-1}}$} & $f_{45\arcsec}$\\
NGC & \multicolumn{3}{c}{(2)} & & \multicolumn{2}{c}{($\rm km\,s^{-1}$)} & $N_c$ & \multicolumn{2}{c}{(6)} & (K) & (\%)\\
\cline{2-4} \cline{6-7} \cline{9-10}
(1)   & ($\arcsec$) & ($\arcsec$) & ($\arcdeg$) & & (3) & (4) & (5) & {\small (mJy/beam)} & (mK)  & (7) & (8) \\
\hline
4192 & 2.4 & 1.9 & 158 && 20.8 & 20.8 & 23 & 18 & 363 & 20.2 & 47\\
4212 & 4.0 & 3.7 & 149 && 20.8 & 10.4 & 22 & 28 & 174 &  6.2 & 76\\
4254 & 3.0 & 2.3 & 148 && 20.8 & 10.4 & 22 & 25 & 333 & 13.3 & 102\\
4303 & 2.8 & 1.9 &  27 && 10.4 & 10.4 & 17 & 21 & 363 & 17.3 & 97\\
4402 & 2.8 & 2.3 & 166 && 20.8 & 10.4 & 20 & 26 & 371 & 14.3 & 106\\
\\
4419 & 3.5 & 2.7 & 159 && 10.4 & 10.4 & 38 & 26 & 253 &  9.7 & 83\\
4501 & 5.6 & 3.7 & 160 && 10.4 & 10.4 & 46 & 17 &  75 &  4.4 & 85\\
4535 & 3.1 & 2.6 & 164 && 10.4 & 10.4 & 23 & 21 & 240 & 11.4 & 208\\
4536 & 2.5 & 1.8 & 173 && 10.4 & 10.4 & 34 & 17 & 347 & 20.4 & 68\\
4548 & 2.6 & 2.0 & 154 && 31.2 & 15.6 & 19 & 22 & 389 & 17.7 & 16\\
\\
4569 & 4.5 & 3.1 & 146 && 10.4 & 10.4 & 37 & 25 & 165 &  6.6 & 111\\
4571 & 3.8 & 2.5 & 154 && ---- & ---- & 0  & 33 & 319 &  9.7 & ---\\
4579 & 4.5 & 3.5 & 152 && 20.8 & 10.4 & -- & 22 & 128 &  5.8 & 31\\
4654 & 5.2 & 3.7 & 149 && 20.8 & 10.4 & 23 & 22 & 105 &  4.8 & 142\\
4689 & 5.2 & 4.2 & 134 && 20.4 & 10.4 & 17 & 20 &  84 &  4.2 & 68\\
\hline
    \end{tabular}
  \end{center}
Notes. ---
Col.(1): Galaxy name.
Col.(2): Major and minor axis sizes and position angle for the
synthesized beam.
Col.(3) and (4): Integrated velocity width of a channel, and
sampling width between channels of the cube.
Col.(5): Number of channels where emission are detected.
Col.(6): Rms noise scaled for a ${\rm 10 \, km\,s^{-1}}$ channel.
Col.(7): Equivalent antenna temprature for ${\rm 1 \, Jy\,beam^{-1}}$.
Col.(8): Fraction of single-dish flux recovered by the aperture
synthesis observations. Single-dish data are from Kenney \& Young (1988).
\end{table}

\clearpage


\begin{table}
  \caption{Adopted Parameters for Analyses}\label{tab:first}
  \begin{center}
    \begin{tabular}{cccc}
\hline \hline
    & \multicolumn{2}{c}{Center Position} &     \\
NGC & \multicolumn{2}{c}{(2)} & Reference \\
\cline{2-3}
(1)  & RA(J2000) & DEC(J2000)   &   (3)    \\
\hline
4192 & 12 13 48.29 & +14 54 01.9 & 1 \\
4212 & 12 15 39.40 & +13 54 04.6 & 2 \\
4254 & 12 18 49.61 & +14 24 59.6 & 1 \\
4303 & 12 21 54.94 & +04 28 25.6 & 1 \\
4402 & 12 26 07.45 & +13 06 44.7 & 1 \\
\\
4419 & 12 26 56.40 & +15 02 50.2 & 1 \\
4501 & 12 31 59.12 & +14 25 13.3 & 1 \\
4535 & 12 34 20.35 & +08 11 52.2 & 1 \\
4536 & 12 34 27.08 & +02 11 17.1 & 1 \\
4548 & 12 35 26.44 & +14 29 47.4 & 1 \\
\\
4569 & 12 36 49.82 & +13 09 45.8 & 3 \\
4579 & 12 37 43.53 & +11 49 05.5 & 4 \\
4654 & 12 43 56.67 & +13 07 36.1 & 1 \\
4689 & 12 50 15.86 & +13 29 27.4 & 1 \\
\hline
    \end{tabular}
  \end{center}
Notes. ---
Col.(1): Galaxy name.
Col.(2): Adopted central position.
Col.(3): Reference of the position:
(1) This study. Dynamical center derived with the AIPS/GAL package;
(2) Cotton et al. 1999;
(3) Sakamoto et al. 1999;
(4) Ho \& Ulvestad 2001.
\end{table}

\begin{table} 
\begin{center}
\caption{Peak CO Brightness Temperatures and Intensities}
\vskip 2mm
\begin{tabular}{cccc}
\hline\hline
\\
NGC & $T_{\rm b, peak}$ & $I_{\rm CO, peak}$ & $I_{\rm CO, peak}\times {\rm cos} i$ \\
  & [K] &  [$\rm K \kms$] & [$\rm K \kms$]\\
\\
\hline
4569 & 3.54 & 586 & 266 \\
4419 & 5.71 & 585 & 229 \\
4536 & 5.87 & 575 & 224 \\
4303 & 6.48 & 436 & 395 \\
4535 & 5.17 & 377 & 275 \\
\\
4192 & 3.26 & 348 &  96 \\
4402 & 4.06 & 224 &  58 \\
4501 & 1.14 & 210 & 111 \\
4548 & 1.46 & 123 &  98 \\
4212 & 1.38 & 115 &  78 \\
\\
4254 & 3.88 & 110 &  97 \\
4579 & 0.78 &  84 &  67 \\
4654 & 1.15 &  44 &  27 \\
4689 & 0.62 &  24 &  21 \\
\hline
\end{tabular}\\
\end{center}
Each value contains a systematic error of about $\pm 15$ \%.
Inclination angles from optical isophotos (Table 1)
are assumed.
\end{table}

\clearpage
\def\deg{$^\circ$}
\def\v{\vskip 4mm}
Figure Captions

\v
PS figures are available at 
http://www.ioa.s.u-tokyo.ac.jp/radio/virgo
\v
\parindent=0pt

Fig. 1. Atlas of the observed Virgo galaxies.
Top top-left panels show DSS second generation blue
optical images, each for $5'\times5'$ area.
$\co$ observations were obtained for the central $1'\times1'$ regions.
Bottom panels show observed CO intensity distributions (left) and
corresponding velocity fields (center).
Position-velocity diagrams along the major axes are shown in the top right
panels.
Parameters of the galaxies and displayed areas are indicated at
the bottom for each galaxy.
Indicated RA and Dec are in J2000.
The north to the top, and the east to the left.
\v

(a){\bf NGC 4192}:

(tl) DSS b-band $5'\times5'$;  $i=74^\circ$; $PA= 155$\deg

(bl) Ico: $1'\times 1'$;  cl=20 $\times$ 1 2,..10,12,,,16,18 K \kmps.

(tr) PVD: $1' \times 3''$, PA 155\deg;  cl=0.1 x (1, 2, ... 10, 12, ... 22) K.

(br) V-field: $1'\times 1'$; cl=-400 to 50, every 50 \kmps.

(b){\bf NGC 4212}:
(tl) DSS b-band $5'\times5'$;  $i=47^\circ$; $PA= 75$\deg

(bl) Ico: $1'\times 1'$;
cl=50 x (0.5, 1, 2, 3, ..., 10) K \kmps.

(tr) PVD: $1' \times 3''$, PA 75\deg;  cl=0.5 x (0.25, 1,2,...10) K.

(br) V-field: $1'\times 1'$; cl=-200 to 40, every 20 \kmps.

\v
(c) {\bf NGC 4254}:

(tl) DSS b-band $5'\times5'$, $i$=42\deg, PA=45\deg

(bl) Ico: $80''\times 80''$;
cl= 10 $\times$(1, 2, ... 12) K \kmps.

(tr) PVD: $1' \times 3''$, PA 45\deg;  cl= 0.132 $\times$(1, 2, ... 12) K.

(br) V-field: $1'\times 1'$; cl= 2300 to  2530, every 20 \kmps.

\v
(d) {\bf NGC 4303}:

(tl) DSS b-band $5'\times5'$, $i$=25\deg, PA=0\deg

(bl) Ico: $1'\times 1'$;
Beam $2''.80\times 1''.90$;
cl= 25 $\times$(1, 2, ... 12) K \kmps.

(tr) PVD: $1' \times 3''$, PA 340\deg;  cl= 0.5$\times$(1, 2, ... 12) K.

(br) V-field: $1'\times 1'$; cl= 1500 to 1600, every 10 \kmps.

\v
(e) {\bf NGC 4402}:

(tl) DSS b-band $5'\times5'$;
$i=75^\circ$; $PA= 90$\deg

(bl) Ico: $80''\times 80''$;
cl=20 x (1, 2, 3, ..., 10) K \kmps.

(tr) PVD: $80' \times 3''$, PA 90\deg;  cl=0.3 $\times$ (1, 2, ... 12) K.

(br) V-field: $80''\times 80''$; cl= 60 to 340, every 20 \kmps.

\v
(f) {\bf NGC 4419}:

(tl) DSS b-band $5'\times5'$;  $i=67^\circ$; $PA= 133$\deg

(bl) Ico: $1'\times 1'$;
cl=50x  (0.4, 0.8, 1.2,  2, 3, ..., 12) K \kmps.

(tr) PVD: $1' \times 5''$, PA 133\deg;  cl= 0.5 x (0.5, 1,2,...10) K.

(br) V-field: $1'\times 1'$; cl= -400 to 0, every 20\kmps.

\v
(g) {\bf NGC 4501}:

(tl) DSS b-band $5'\times5'$; $i=58^\circ$; $PA= 140$\deg

(bl) Ico: $80''\times 80''$;
cl = 10 $\times$ (0.5, 1.0, 1.5, 2, 3, .., 10, 12, ..., 20) K \kmps.

(tr) PVD: $1' \times 10''$, PA 140\deg;  cl= 0.1 x (0.5, 1,2,...10) K.

(br) V-field: $80''\times 80''$; cl= 200 to 500, every 50 \kmps.

\v
(h) {\bf NGC 4535}:

(tl) DSS b-band $5'\times5'$, $i$=43\deg, PA=0\deg

(bl) Ico: $1'\times 1'$;
Beam $2''.80\times 1''.90$;
cl= 10 $\times$(1, 2, 4, 6, ... 10, 15, 20, 25) K \kmps.

(tr) PVD: $1' \times 10''$, PA 0\deg;  cl= 0.5$\times$(1, 2, ... 12) K.

(br) V-field: $1'\times 1'$; cl= 1800 to 2100, every 10 \kmps.

\v
(i) {\bf NGC 4536}:

(tl) DSS b-band $5'\times5'$;  $i=67^\circ$; $PA= 116$\deg

(bl) Ico: $1'\times 1'$;
cl=20 $\times$ (1, 2, 4, 6, ... 10, 15, 20, 25, 30) K \kmps.

(tr) PVD: $1' \times 5''$, PA 116\deg;  cl=0.5 x (0.25, 1,2,...10) K.

(br) V-field: $1'\times 1'$; cl=1600 to 2000, every 50 \kmps.

\v
(j) {\bf NGC 4548}:
(tl) DSS b-band $5'\times5'$;  $i=37^\circ$; $PA= 150$\deg

(bl) Ico: $1'\times 1'$;
cl=10 x (1, 2, 3, ..., 12) K \kmps.

(tr) PVD: $1' \times 5''$, PA 150\deg;  cl=0.15 x (1,2,...10) K.

(br) V-field: $1'\times 1'$; cl=300 to 700, every 50 \kmps.

\v
(k) {\bf NGC 4569}:

(tl) DSS b-band $5'\times5'$;  $i=63^\circ$; $PA= 23$\deg

(bl) Ico: $1'\times 1'$;
cl= 50$\times$(0.25, 0.5, 1, 2, ... 12) K \kmps.

(tr) PVD: $1' \times 5''$, PA 160\deg;
cl= 0.2$\times$(1, 2, ..... 15) K

(br) V-field: $1'\times 1'$;
 cl= -380 to 0, every 20 km/s.

\v
(l) {\bf NGC 4579}:

(tl) DSS b-band $5'\times5'$;  $i=37^\circ$; $PA= 60$\deg

(bl) Ico: $1'\times 1'$;
cl=20 x (1, 2, 3, ..., 7) K \kmps.

(tr) PVD: $1' \times 5''$, PA 90\deg;  cl=0.25 x (1,2, ..., 6) K.

(br) V-field: $1'\times 1'$; cl=1320 to 1600, every 20 \kmps.

\v
(m) {\bf NGC 4654}:

(tl) DSS b-band $5'\times5'$; $i=51^\circ$; $PA= 128$\deg

(bl) Ico: $1'\times 1'$;
cl=5 x  (1, 2, ... 10) K \kmps.

(tr) PVD: $1' \times 3''$, PA 128\deg;
cl= 0.1 x (1, 2,... 10) K.

(br) V-field: $1'\times 1'$; cl= 960 to 1120, every 20\kmps.

\v
(n) {\bf NGC 4689}:

(tl) DSS b-band $5'\times5'$;  $i=30^\circ$; $PA= 160$\deg

(bl) Ico: $1'\times 1'$;
cl= 2.5 x (1, 2, 3, ..., 10) K \kmps.

(tr) PVD: $1' \times 5''$, PA 160\deg;
cl=0.05  x (1,2,...10) K.

(br) V-field: $1'\times 1'$; cl= 1500  to 1700, every 20 \kmps.

\v
Fig. 2. Integrated $\co$-intensity maps of the
observed galaxies in the same angular scale.
The image sizes are $1.'0 \times 1.'0$, or 4.68 kpc $\times 4.68$ kpc
for an assumed distance of 16.1 Mpc.
The contours are drawn at 5, 10, 20, 40, 80, 160 K $\kms$.

\v
Fig. 3. CO line velocity fields of the observed galaxies in the same
angular scale corresponding to figure 2.
Contours are drawn every 20 $\kms$ relative to the systemic velocity,
which is expressed by white thick contours.
Darker coding represents redshift, and white for blueshift.

\v
Fig. 4. Sky plot of $\ico$ on the Virgo Cluster area.
Each map is enlarged by 50 times the real angular size.
The position of M87 is marked by a cross.

\v
Fig. 5. Radial profiles of the face-on CO intensity obtained
by ellipse fitting.
The plotted radius is $40''$.
The primary-beam attenuation has been corrected.

\v
Fig. 6. CO intensity distributions in the central $20''\times 20''$ regions
of the  "central/single-peak" galaxies. Contour levels are
20 $\times$ (1, 2, ..., 10, 12, ... 20, 25, ... 40) K $\kms$.

\newpage
\def\vlsr{$V_{\rm lsr}$}
\noindent
{\bf \Large Appendix}

We show the channel maps of individual galaxies in Figure A1, particularly
to confirm that no significant continuum emission has been detected.

\v
Fig. A1. Channel maps of the $^{12}$CO ($J=1-0$) line emission of
the Virgo galaxies. Intensity scale is in Kelvin
of brightness temperature.
Contours are drawn at $2^n$ times the lowest-contour value
($n=1, 2, 3, .....$). The lowest contour level (cl) is
indicated for individual galaxies.

\v

NGC 4192: Lowest cl= 0.5 K.

NGC 4212: Lowest cl = 0.25 K.

NGC 4254: Lowest cl = 0.5 K.

NGC 4303: Lowest cl = 1.0 K.

NGC 4402: Lowest cl = 0.5 K.

NGC 4419: Lowest cl = 0.5 K.

NGC 4501: Lowest cl = 0.20 K.

NGC 4535: Lowest cl = 0.5 K.

NGC 4536: Lowest cl = 0.5 K.

NGC 4548: Lowest cl = 0.25 K.

NGC 4569: Lowest cl = 0.5 K.

NGC 4579: Lowest cl  = 1.0 K.

NGC 4654: Lowest cl = 0.5 K.

NGC 4689: Lowest cl = 0.125 K.

\end{document}